\documentclass[twocolumn,showpacs,floatfix,pre]{revtex4-1}

\usepackage{graphicx}
\usepackage{graphicx}
\usepackage{dcolumn}
\usepackage{epsf}
\usepackage{amsmath}
\usepackage{amssymb}
\usepackage{bm}

\newcommand{\pdf}{{\it pdf\,}}

\begin{document}


\title{Atomic Level Green-Kubo Stress Correlation Function for a
Model Crystal:\\ An Insight  into Molecular Dynamics Results on a Model Liquid.}


\author{V.A. Levashov}
\affiliation{Department of Physics and Astronomy, University of
Tennessee, Knoxville, TN 37996, USA.}



\begin{abstract}
In order to get insight into the connection between the vibrational dynamics 
and the {\it atomic level} Green-Kubo stress correlation function in liquids 
we consider this connection in a model crystal instead.
Of course, vibrational dynamics in liquids and crystals are quite different and 
it is not expected that the results obtained on a model crystal 
should be valid for liquids.  
However, these considerations provide  a benchmark to which the results of 
the previous molecular dynamics simulations can be compared.
Thus, assuming that vibrations are plane waves, we derive analytical expressions
for the atomic level stress correlation functions in the classical limit and analyze them.
These results provide, in particular, a recipe for analysis of the atomic level stress correlation functions
in Fourier space and extraction of the wavevector and 
frequency dependent information.  
We also evaluate the energies of the atomic level stresses.
Obtained energies are significantly smaller than the energies that were obtained in MD
simulations of liquids previously. 
This result suggests that the average energies of the atomic level stresses in liquids and glasses
are largely determined by the structural disorder.
We discuss this result in the context of equipartition of the atomic level stress energies. 
Analysis of the previously published data suggests that it is possible to speak about
configurational and vibrational contributions to the average energies of the atomic level
stresses in a glass state. However, this separation in a liquid state is problematic.
We also consider peak broadening in the pair distribution function with increase of distance.
We find that peak broadening (by $\approx 40\%$) occurs due 
to the transverse vibrational modes, while contribution from the longitudinal modes does not 
change with distance. 
Finally, we introduce and consider atomic level transverse current correlation function.
\end{abstract}



\maketitle

\section{Introduction}\label{sec:intro}

 
In order to understand abrupt increase in viscosity of liquids approaching 
the glass transition, it is necessary to understand well the nature of viscosity itself.
This understanding, however, is still limited 
\cite{HansenJP20061,EvansDJ19901,Boon19911,Hetero20131,Berthier2011,
Tanaka20091,Tanaka2011E,Levashov20111,Levashov2013,Iwashita2013}.

Computer simulations has proved to be an important tool in addressing properties of
supercooled liquids 
\cite{HansenJP20061,EvansDJ19901,Boon19911,Hetero20131,Berthier2011}.
One standard approach to calculate viscosity in 
computer simulations is based on the Green-Kubo expression that relates
viscosity to the integral of the macroscopic stress correlation function 
\cite{HansenJP20061,EvansDJ19901,Boon19911,Green1954,Kubo1957,Helfand1960,EvansDJ19811,Hoheisel19881}. 

Properties of the stress correlation function have been extensively studied previously
from {\it a macroscopic} perspective \cite{HansenJP20061,EvansDJ19901,Boon19911,Hoheisel19881}. 
There have been significantly fewer studies that
tried to address how behavior of the system at the atomic level translates into
the macroscopic behavior of the stress correlation function 
\cite{Woodcock19911,Woodcock20062,Stassen19951,Stassen19952,Levashov20111,Levashov2013,
Iwashita2013}. 

The situation is similar with a closely related but somewhat different approach, 
i.e., the approach based on considerations of the transverse current correlation function 
\cite{HansenJP20061,EvansDJ19901,Boon19911,Tanaka20091,Tanaka2011E,Tanaka20081,Mizuno2013,Mountain19821}.
Studies of vibrational dynamics in disordered media with the transverse current correlation 
function are very common and several important results were obtained with it 
relatively recently \cite{Tanaka20091,Tanaka2011E,Tanaka20081}.
However, in all these studies the transverse current correlation function
is treated as a {\it macroscopic} quantity. Thus the relations between the 
atomic level processes and the macroscopic behavior of the transverse current
correlation function remain obscure \cite{Tanaka20091,Tanaka2011E}.  

We previously studied {\it atomic level structure} of the macroscopic 
Green-Kubo stress correlation function by decomposing it into 
correlation functions between the atomic 
level stresses \cite{Levashov20111,Levashov2013}. 
The approach represents further development of preceding 
works \cite{Hoheisel19881,
Woodcock19911,Woodcock20062,Stassen19951,Stassen19952}. 
Our data clearly show presence of stress waves in the 
atomic level stress correlation function and that the stress waves contribute to viscosity 
\cite{Levashov20111,Levashov2013}.
However, it was not previously discussed {\it how} stress waves and their properties 
translate into the observed atomic level stress correlation functions. 
It is difficult to address this issue in liquids, even qualitatively, 
as vibrational and configurational dynamics in liquids are mixed 
\cite{Stillinger20131,HeuerA20081,KeyesT19971}. 
Moreover, vibrational and configurational dynamics
in disordered media are puzzles by themselves 
\cite{Stillinger20131,HeuerA20081,Zwanzig19651,KeyesT19971,
Taraskin2000,Taraskin2002,Scopigno20071,
Tanaka20081,Mizuno2013,Keys2011,
Iwashita2013,Kob2006,Frenkel1947,Trachenko2009,
Bolmatov2012,Bolmatov2013}.

On the other hand, as it appears from the review of the previous
literature, the details of the connection between vibrational dynamics
and {\it the atomic level} stress correlation function were not addressed
previously even for those systems for which it could be done relatively easily, 
i.e., for the crystals. 
Applicability of results obtained from crystal models to liquids, in general, 
is not expected and should be considered with caution.
However, it has been demonstrated
that parallels between liquid and solid states can be 
useful \cite{Frenkel1947,Trachenko2009,Bolmatov2012,Bolmatov2013}.

Thus, in order to gain at least some qualitative or semiquantitative insight into the 
connection between the vibrational dynamics of a model liquid and
the atomic level stress correlation functions observed in 
MD simulations \cite{Levashov20111,Levashov2013}, 
we examine a crystal-like model  in which vibrations
are represented by plane waves.
Considerations in this paper represent further developments and more detailed
discussions of some ideas and a model first presented in Ref.\cite{Egami19821}.

Another goal of this paper is to develop a framework 
for analysis in Fourier space of the MD data from a model 
liquid \cite{Levashov20111,Levashov2013}. 
This analysis is presented in Ref.\cite{Levashov20141}.
It relies on the results presented in this paper. 

To make derivations of the expressions for the stress correlation functions clearer
it is useful to address several other issues.
In particular, we calculate atomic level stress energies. 
In this context we discuss the data from previously published MD simulations 
on liquids and glasses \cite{Egami19821,Chen19881,Levashov2008B,Levashov2008E}.

In the framework of the model it is easy to evaluate the peak broadening in the pair 
distribution function with increasing distance. Calculations show
that the peak broadening (by $\approx 40\%$) occurs 
because of the transverse waves, while the contribution from the longitudinal 
waves only weakly depends on distance.

Finally we briefly discuss {\it the atomic level transverse current correlation function}
and argue that it is possible to study its behavior in MD simulations in a way 
which we previously applied to the atomic level stress correlation function.

The paper is organized as follows. 
In section \ref{sec:model} we describe the model.
Section \ref{sec:derivations} is focused
on derivations and analysis of the obtained
results. In section \ref{sec:discussion}
we discuss obtained results in the broader
context of some results obtained previously. 

\section{The model \label{sec:model}}

We consider a single component system and 
assume that different atoms have identical environments.
In particular, we assume that every atom interacts harmonically with $N_c$ nearest neighbors. 
We also assume that distribution of these neighbors is 
spherically symmetric and that their equilibrium distance from the central atom
is $a$.
Finally we assume that vibrational motion in the system
is described by plane waves.

\subsection{Continuous spherical approximation}

In the following derivations we will usually perform summation for every 
atom $n$ over its nearest neighbors $m$. 
In performing these summations we will utilize a continuous spherical approximation.
Thus we will change summation over $m$ into 
the integration over the spherical angles:
\begin{eqnarray}
\sum_m f(\theta_m, \phi_m ) \rightarrow \frac{N_c}{4\pi}\int f(\theta ,\phi)\sin(\theta)d\theta d\phi
\;\;.\;\;\;\;\;\;\;\;
\label{eq;sphericalapprox01}
\end{eqnarray}

\subsection{Debye's Model}

In order to estimate various quantities to which many different waves
contribute we will assume that different waves contribute independently.
We will also utilize Debye's model, i.e., we will change summation
over different waves into the integration over the wavevector: 
\begin{eqnarray}
\frac{dN}{N}=\left(\frac{a}{2\pi}\right)^3\,4\pi\,q^2\,dq\;\;,\;\;\;\;Q_{max} =\left(\frac{\pi}{a}\right)\left(\frac{6}{\pi}\right)^{1/3}\;\;,\;\;\;\;\;\;\;\;
\label{eq;debye01}
\end{eqnarray}
where $N$ is the total number of atoms in the system 
and also the total number of vibrational states for 
one polarization.
$dN$ is the number of states in the interval $dq$,
and $Q_{max}$ is the maximum value of the wavevector.
Equations in (\ref{eq;debye01}) are written for one 
particular polarization of the waves. We will assume
further, as usual, that there are one longitudinal and two 
transverse polarizations.

The value of $Q_{max}$ and the value of the prefactor $\left(a/(2\pi)\right)^3$ in 
(\ref{eq;debye01}) are connected by the normalization condition. In principle,
one can assume different values of $Q_{max}$ for different 
polarizations of the waves.
We will not elaborate on this issue further.
 
\subsection{Long wavelength approximation}

In the following we will sometimes assume that:
\begin{eqnarray}
\sin\left(\bm{q}\bm{a}_{nm}\right)\approx \left(\bm{q}\bm{a}_{nm}\right)
\;\;,\;\;\;
\cos\left(\bm{q}\bm{a}_{nm}\right)\approx 1\;\;.\;\;\;\;\;\;\;\;
\label{eq;LWapprox01}
\end{eqnarray}
Equations (\ref{eq;LWapprox01}) are correct if the wavelength 
of the wave is much larger than the interatomic distance $a\equiv |\bm{a}_{nm}|$.
Usually we will give the results obtained without long wavelength approximation and
then, for comparison, the results obtained with long wavelength approximation. 

\section{Derivations \label{sec:derivations}}

\subsection{Potential energy of an atom due to a plane wave}

Let us assume that $\bm{r}_n^o$ is the equilibrium position of the particle $n$ 
and $\bm{u}_n$ is the displacement of the particle $n$ from equilibrium. 
Then 
$\bm{r}_n = \bm{r}_n^o + \bm{u}_n$,
$r_{nm} = \left|\bm{r}_m - \bm{r}_n\right|$,
$\bm{a}_{nm}=a\bm{\hat{a}}_{nm}=\left(\bm{r}_m^o-\bm{r}_n^o\right)$,
$\bm{u}_{nm}=\bm{u}_m-\bm{u}_n$.
With these notations potential energy for the nearest neighbor atoms $n$ and $m$ 
in the harmonic approximation is given by:
\begin{eqnarray}
U_{nm}=\frac{k\left(r_{nm}-r_{nm}^o\right)^2}{2} \approx 
\frac{k}{2}\left(\bm{\hat{a}}_{nm}\bm{u}_{nm}\right)^2\;\;\;\;.
\label{eq;poten3dx1}
\end{eqnarray}

The solutions for particle displacements
in classical harmonic crystals are plane waves. 
For a particular wave:
\begin{eqnarray}
&&\bm{u}_n(\bm{q}) =u_q\bm{\hat{e}}_q\,Re \left\{\chi_n(\bm{q})\right\}\label{eq;poten3dx21}\;\;\;,\;\;\;\\
&&\chi_n(\bm{q}) =
\exp\left[-i\left(\omega_{\bm{q}} t - \bm{q}\bm{r}_n +\phi_{\bm{q}}\right)\right]\;\;\;,\;\;\;
\label{eq;poten3dx22}
\end{eqnarray}
where $u_q$ (real scalar) is the amplitude of the wave 
and $\bm{\hat{e}}_q$ (real vector)
is its polarization vector.

From (\ref{eq;poten3dx21},\ref{eq;poten3dx22}) we get:
\begin{eqnarray}
\bm{u}_{nm}(\bm{q}) = u_q\bm{\hat{e}}_q
Re\left\{\chi_n(\bm{q})\left[\exp\left(i\bm{q}\bm{a}_{nm}\right) - 1\right]\right\}\;\;.
\label{eq;unm3plane1}
\end{eqnarray}

It is straightforward to show from (\ref{eq;poten3dx1},\ref{eq;unm3plane1}) 
that time average of the potential energy of the atom $n$ due to a particular wave is:
\begin{eqnarray}
\left<U_{n}\right>_t \approx \left(\frac{1}{2}\right)\,ku_q^2\sum_m\left(\bm{\hat{a}}_{nm}\bm{\hat{e}}_q\right)^2 
\sin^2\left(\frac{\bm{q}\bm{a}_{nm}}{2}\right)\;\;,
\label{eq;poten3dx3}
\end{eqnarray}
where we introduced factor $1/2$ to take into account that 
half of the elastic energy belongs to the atom $n$, 
while another half to the atom $m$.

\subsection{Force on an atom and dispersion relations}

It follows from (\ref{eq;poten3dx1}) that the force on the atom $n$ 
due to its interaction with the atom $m$ is:
\begin{eqnarray}
f_{nm}^{\alpha} = -\frac{\partial U_{nm}}{\partial u_n^{\alpha}}=k(\bm{\hat{a}}_{nm}\bm{u}_{nm})\hat{a}_{nm}^{\alpha}\;\;. 
\label{eq;force01}
\end{eqnarray}
Using the expression (\ref{eq;unm3plane1}) for $\bm{u}_{nm}(\bm{q})$ in (\ref{eq;force01}) for the total
force on the atom $n$ we get:
\begin{eqnarray}
f_{n}^{\alpha} = && \sum_m (ku_q)\,\left(\bm{\hat{a}}_{nm}\bm{\hat{e}}_{\bm{q}}\right)\,
a_{nm}^{\alpha}\cdot\label{eq;force02}\\
&&
Re\left\{\chi_n(\bm{q})\left[\exp\left(i\bm{q}\bm{a}_{nm}\right) - 1\right]\right\}\;\;\;\;.\;\;
\nonumber
\end{eqnarray}
Let us further suppose that we consider crystal lattices with central symmetry.
Then for every neighbor $m$ there is another neighbor $m'$ 
such that $\bm{a}_{nm'}=-\bm{a}_{nm}$. 
This assumption should be true in the continuous spherical approximation.
Taking this into account we can rewrite (\ref{eq;force02}) as:
\begin{eqnarray}
f_{n}^{\alpha} = && \sum_m \left(ku_q\right)\left(\bm{\hat{a}}_{nm}\bm{\hat{e}}_{\bm{q}}\right)
a_{nm}^{\alpha}\cdot \label{eq;force03}\\
&&\left[\cos\left(\bm{q}\bm{a}_{nm}\right) - 1\right]\cdot 
\cos\left(\omega_{\bm{q}} t - \bm{q}\bm{r}_n +\phi_{\bm{q}}\right)\;\;\;\;.\;\;\;\;
\nonumber
\end{eqnarray}
From (\ref{eq;poten3dx21},\ref{eq;poten3dx22},\ref{eq;force03}) 
and Newton's second law we get:
\begin{eqnarray}
\hat{e}_{\bm{q}}^{\alpha}\omega_{L,T}^2(\bm{q})=\left(\frac{2k}{M}\right)\sum_m
\left(\bm{\hat{a}}_{nm}\bm{\hat{e}}_{\bm{q}}\right)
\sin^2\left(\frac{\bm{q}\bm{a}_{nm}}{2}\right) a_{nm}^{\alpha}
\;\;,\;\;\;\;\;\;\;
\label{eq;dispersion3D01}
\end{eqnarray}
where $M$ is the particle's mass.
Indexes $L$ and $T$ label longitudinal and transverse polarizations. 
Multiplication of both sides of (\ref{eq;dispersion3D01}) on $\hat{e}^{\alpha}_{\bm{q}}$
with the following summation over $\alpha$ leads to:
\begin{eqnarray}
\omega_{L,T}^2(\bm{q})=\left(\frac{2k}{M}\right)\sum_m
\left(\bm{\hat{a}}_{nm}\bm{\hat{e}}_{\bm{q}}\right)^2
\sin^2\left(\frac{\bm{q}\bm{a}_{nm}}{2}\right)
.\;\;\;\;
\label{eq;dispersion3D02}
\end{eqnarray}
Expression (\ref{eq;dispersion3D02}) is very similar to expression
(\ref{eq;poten3dx3}). 
This, of course, is not an occasion as the average potential energy of a site
(\ref{eq;poten3dx3}) due to a wave with the 
amplitude $u_{\bm{q}}$ should be equal to $M\omega^2_{\bm{q}} u^2_{\bm{q}}/4$.
\begin{figure}
\begin{center}
\includegraphics[angle=0,width=3.3in]{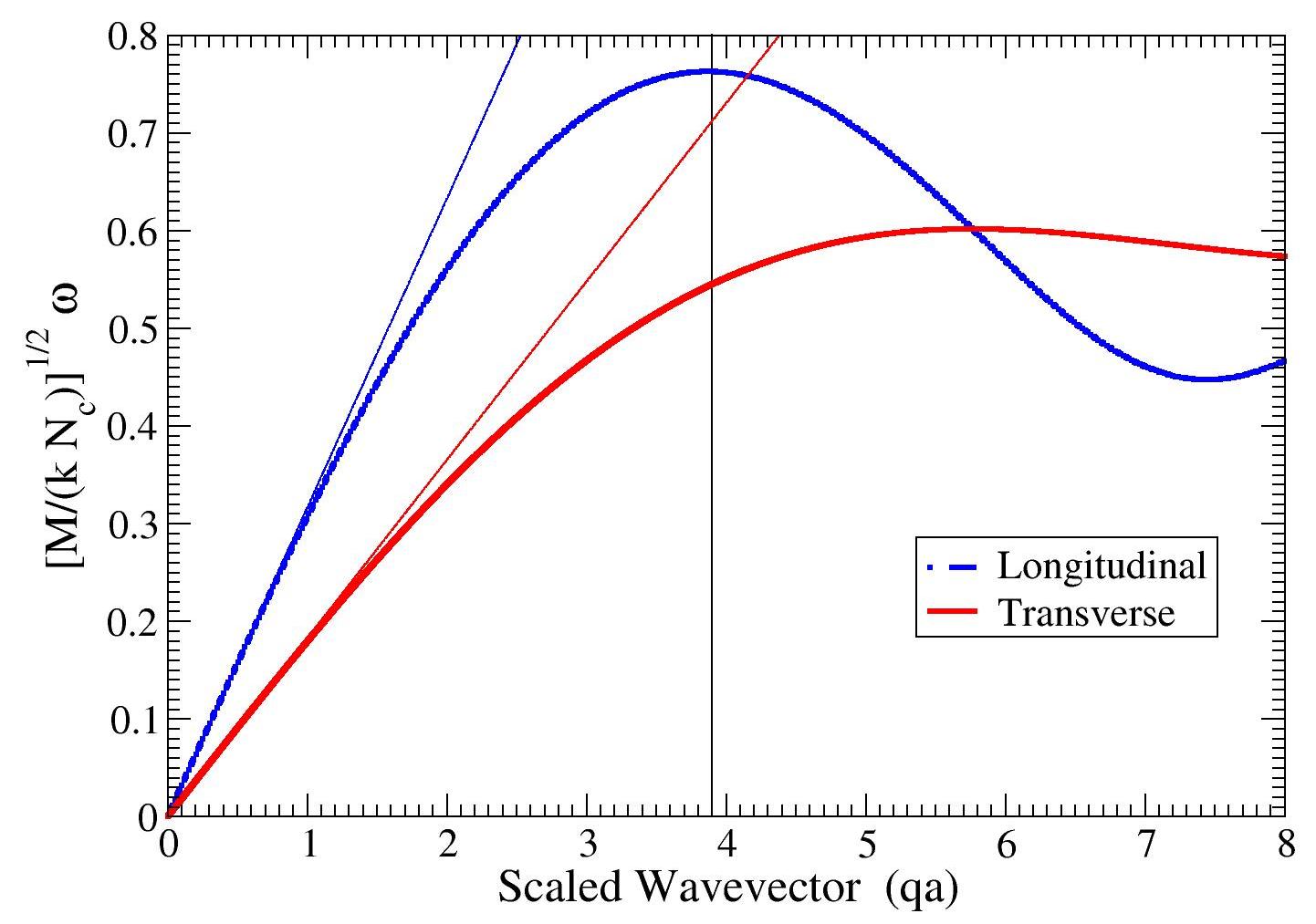}
\caption{Dispersion curves for the longitudinal and transverse waves in the 
continuous spherical approximation. Thick lines show the results obtained
without the long wavelength approximation. Thin lines show the results obtained
with the long wavelength approximation. 
Note that $\left(Q_{max}a\right) \cong 3.9$
}\label{fig:dispersion02}
\end{center}
\end{figure}

In the continuous spherical approximation (\ref{eq;dispersion3D02}) should
not depend on the direction of $\bm{q}$ and
an analytical expression for (\ref{eq;dispersion3D02}) 
could be obtained for the longitudinal and transverse waves.
For a longitudinal wave it is sufficient to assume that
$(\bm{q} \parallel \bm{\hat{z}})$ and $(\bm{\hat{e}_q} \parallel \bm{\hat{z}})$.
For a transverse wave it is sufficient to assume that
$(\bm{q} \parallel \bm{\hat{z}})$ and $(\bm{\hat{e}_q} \parallel \bm{\hat{x}})$.
Thus we can rewrite (\ref{eq;dispersion3D02}) as:
\begin{eqnarray}
\omega_{L,T}^2(q)=\omega_o^2 D_{L,T}(qa)\equiv\frac{2}{4\pi}
\int f_{L,T}(\xi,\theta,\phi)d\Omega\;\;\;,\;\;\;\;
\label{eq;dispersion3D03}
\end{eqnarray}
where 
\begin{eqnarray}
\omega_o^2 \equiv \left(\frac{k}{M}\right)N_c\;\;\;,\;\;\;\;\;\;\xi\equiv \frac{qa}{2}\;\;\;.\;\;
\label{eq;xi}
\end{eqnarray}
For longitudinal and transverse waves:
\begin{eqnarray}
&&f_L(\xi,\theta,\phi)=\cos^2(\theta)\,\sin^2\left(\xi\cos(\theta)\right)\;\;\;,\;\;\nonumber\\
&&f_T(qa,\theta,\phi)=\sin^2(\theta)\cos^2(\phi)\,
\sin^2\left(\xi\cos(\theta)\right)\;\;\;.\;\;\;\;\;\nonumber
\end{eqnarray}
Integrations over the spherical angles using 
the Maple(\textsuperscript{TM}) program \cite{Maple5} lead to:
\begin{eqnarray}
&&D_L(qa)= \left[L_1(\xi)+L_2(\xi)\right]/\left[6\xi^3\right]\;\;\;,\;\;\;\\
&&D_T(qa)= \left[T_1(\xi)+T_2(\xi)\right]/\left[12\xi^3\right]\;\;\;,\;\;\;
\label{eq;poten3dx3sin2}
\end{eqnarray}  
where
\begin{eqnarray}
&&L_1(\xi)=- 6\,\xi^{2}\cos(\xi)\,\sin(\xi) + 2\,\xi^{3}
 - 6\,\xi\,\cos^2(\xi)\;\;\;,\nonumber\\
&&L_2(\xi) = 3\,\cos(\xi)\,\sin(\xi) + 3\,\xi\;\;\;,\nonumber\\
&&T_1(\xi)=4\xi^{3} + 6\,\xi\,\cos^2(\xi)\nonumber\\
&&T_2(\xi) = -3\,\cos(\xi)\,\sin(\xi) - 3\,\xi\;\;\;.
\label{eq;poten3dx3sin3}
\end{eqnarray}  
The dependencies $\sqrt{D_L(qa)}$ and $\sqrt{D_T(qa)}$ on $qa$, 
i.e., the dispersion relations, are plotted in Fig.\ref{fig:dispersion02}.

The dispersion relations for the longitudinal and transverse waves could also be calculated 
in the long wavelength ({\it lw}) approximation:
\begin{eqnarray}
&&\omega_{lw}^2(L,q) = \frac{\omega_o^2}{10}(qa)^2,\;\;\;\;\;
\omega_{lw}^2(T,q) = \frac{\omega_o^2}{30}(qa)^2.\;\;\;\;\; 
\label{eq;dispersionlw}
\end{eqnarray} 
Thus in the long wavelength approximation speeds of the longitudinal waves are
$\sqrt{3}$ times larger than the speeds of the transverse waves.

\subsection{Equipartition and mean square displacements}
 
If we will assume that equipartition holds for our spherical approximation 
then the average potential energy of every wave should be equal to $k_b T/2$.
Thus we should have:
\begin{eqnarray}
&&\frac{M\omega_{L,T}^2(q) u_{L,T}^2(q)}{4}=\frac{1}{2}\frac{k_b T}{N}\;\;,\nonumber\\
&&u_{L,T}^2(q)=2\left(\frac{k_bT}{kN_c}\right)\left(\frac{1}{D_{L,T}(qa)}\right)\frac{1}{N}\;\;,\;\;\;\;\;\;\;
\label{eq;equip3D01}
\end{eqnarray}
where $u_{L,T}^2(q)$ is the average square amplitude of the 
longitudinal or transverse waves with the magnitude of the wave vector $q$. 
Thus the squares of the amplitudes are inversely proportional to the
dispersion curves shown in the Fig.\ref{fig:dispersion02}. 
Note that wave's amplitudes diverge for small wavevectors.  

\subsection{Mean square displacements due to all waves}

In order to find the mean square displacements due to all waves, 
assuming that all of them are independent, we have to take half (since ($<u_n^2>=(1/2)u_q^2$)
of (\ref{eq;equip3D01}) and integrate it over all $q$ using (\ref{eq;debye01}).

In this way for the mean square displacements due to all longitudinal waves 
and both polarizations of all transverse waves we get:
\begin{eqnarray}
&&<u^2(L,T)>=\left(\frac{k_bT}{kN_c}\right)\gamma(L,T)\;\;,\;\;\;
\label{eq;u2longAll01}\\
&&\gamma(L)=2.8160\;\;,\;\;\;\;\;\gamma(T)=13.3615\;\;.\;\;\;\nonumber
\end{eqnarray}
Note that $<u^2(T)>$ is significantly larger than
$<u^2(L)>$.

It is simpler to evaluate the values of the mean square displacements 
in the long wavelength approximation. In this case we get:
\begin{eqnarray}
\gamma_{lw}(L)=1.9746\;\;,\;\;\;\gamma_{lw}(T)=11.8478\
\label{eq;u2longWaveGamma01}
\end{eqnarray}
Note that the values of the coefficients in the long wavelength approximation
are smaller than without the long wavelength approximations. 
This is consistent with (\ref{eq;equip3D01}), as 
the values of the frequencies are always larger in the long wavelength 
approximation.
 
\subsection{The widths of peaks in the pair distribution function}

Atoms located close to each other in the lattice should exhibit a certain
degree of coherence in their motion. 
Because of this peaks in the
pair distribution function at small distances should be narrower than at large distances.
The dependence of the peak's widths on distance was investigated previously using
a detailed model and evolved simulations \cite{Chung1997,Chung1999}. 
It the frame of our model we can provide
a simple evaluation of the size of the effect.

The average square of the peak width in the pair distribution function 
is determined by \cite{Chung1997,Chung1999}:
\begin{eqnarray}
\left<\left(\Delta r_{nm}\right)^2\right>\cong
\left<\left(\bm{\hat{r}}_{nm}^o \bm{u}_{nm}\right)^2\right>\;\;\;.\;\;\;
\label{eq;pdf3D01}
\end{eqnarray}
In (\ref{eq;pdf3D01}) the notation $\left<...\right>$ is used for
the time and spherical averages. 
Expression for $\left<\left(\Delta r_{nm}\right)^2\right>$ 
is completely analogous to the expression
(\ref{eq;poten3dx1}), 
but with $\bm{r}_{nm}^o$ instead of $\bm{a}_{nm}$.

\begin{figure}
\begin{center}
\includegraphics[angle=0,width=3.3in]{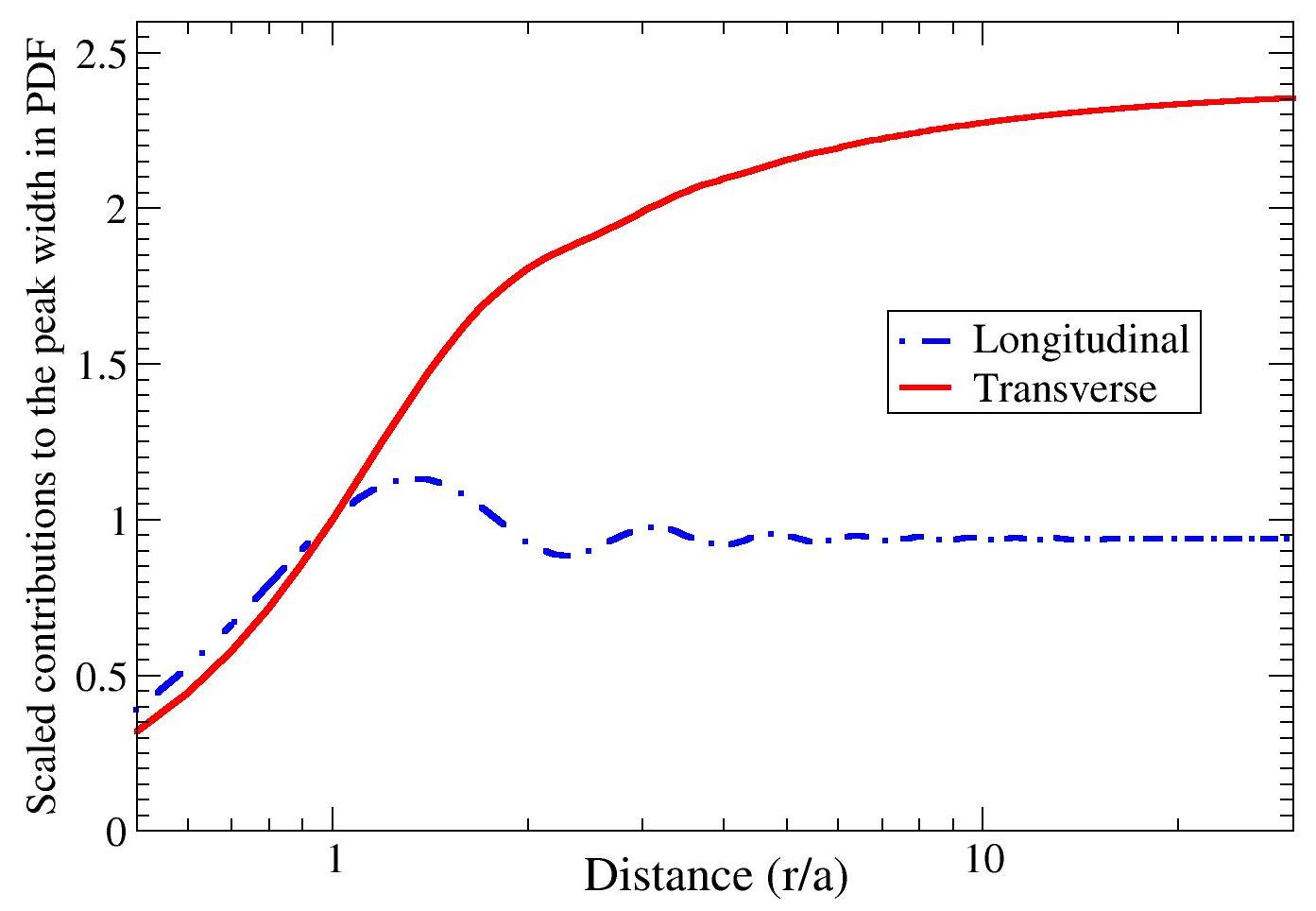}
\caption{The value of the ratio of sums in (\ref{eq;pdf3D03}) 
integrated over all wavevectors as a function of $r$. 
The blue curve represents contributions from all longitudinal waves.
The red curve represents contributions from one polarization of all transverse 
waves.
}\label{fig:uij3D01}
\end{center}
\end{figure}

In analogy with (\ref{eq;poten3dx1},\ref{eq;poten3dx3}) and
using the expression (\ref{eq;equip3D01}) for $u_{\bm{q}}^2$ we get:
\begin{eqnarray}
\frac{k\left<\left(\Delta r_{nm}(\bm{q}\right)^2\right>}{2 k_b T} \cong
\frac{\sum_m\left(\bm{\hat{r}}_{nm}\bm{\hat{e}}_{\bm{q}}\right)^2 
\sin^2\left(\frac{\bm{q}\bm{r}_{nm}}{2}\right)}
{\sum_m\left(\bm{\hat{a}}_{nm}\bm{\hat{e}}_{\bm{q}}\right)^2 
\sin^2\left(\frac{\bm{q}\bm{a}_{nm}}{2}\right)}\;\;\;\;\;\;
\label{eq;pdf3D03}
\end{eqnarray}
In order to estimate the peak width due to all waves it is necessary
to integrate the numerator and denominator of (\ref{eq;pdf3D03}) over the 
spherical angles and then their ratio over all $q$ using (\ref{eq;debye01}).
The results of these integrations 
(assuming that $N_c =1$) for all 
longitudinal waves and one polarization 
of all transverse waves are shown in 
Fig.\ref{fig:uij3D01}.

For large $r_{nm}$ motions of the atoms 
$n$ and $m$ should be uncorrelated. 
It is straightforward to show from (\ref{eq;pdf3D01}) that
if atoms $n$ and $m$ vibrate independently then
$<(\Delta r_{nm})^2> = (2/3) <(u_n)^2>$.
This should be the large $r_{nm}$ limit of the peak's width.
In order to get this limit from the curves in
Fig.\ref{fig:uij3D01} it is necessary to multiply the limiting value
by 2 for the longitudinal waves (the prefactor in \ref{eq;pdf3D03}) and by
4 for the transverse waves (the prefactor and two polarizations). 
Then the results can be compared with (\ref{eq;u2longAll01}).

It is interesting that convergence to the final
value for the transverse waves is slower than for the longitudinal waves. 
Note also that contribution
to the peak width from the shear waves increases by more than twice 
as the distance increases.
There is essentially no change in the peak's widths with distance due to the 
longitudinal waves.

\subsection{Atomic level stress elements}

Similarly to the previous definitions \cite{Egami19821,Chen19881,Levashov2008B},  
we define the $\alpha\beta$-component 
of the local atomic stress element on a particle $n$ as:
\begin{eqnarray}
s_n^{\alpha\beta}=\frac{1}{2}\sum_{m \neq n} f_{nm}^{\alpha}  r_{nm}^{\beta}\;\;,
\label{eq;stress3elem1}
\end{eqnarray}
where, $f_{nm}^{\alpha}$ is the $\alpha$-component of the force on the particle $n$ caused by 
the interaction with the particle $m$ and $r_{nm}^{\beta}$ is  
the $\beta$-component of the radius vector 
from the particle $n$ to the particle $m$. 
The sign in (\ref{eq;stress3elem1}) was chosen in such a way 
that an atom under compression will have a negative stress/pressure.

If there are interactions between the nearest neighbors only, we can rewrite
(\ref{eq;stress3elem1}) using (\ref{eq;force01}) as:
\begin{eqnarray}
s_n^{\alpha\beta}=\frac{\left(ka\right)}{2}\sum_{m \neq n} (\bm{u}_{nm} \bm{\hat{a}}_{nm})
\hat{a}_{nm}^{\alpha}\hat{a}_{nm}^{\beta}\;\;,
\label{eq;stress3elem2}
\end{eqnarray}
 
Using (\ref{eq;unm3plane1}) in (\ref{eq;stress3elem2})  for the complex stress we obtain:
\begin{eqnarray}
&&s_{n}^{\alpha\beta}(\bm{q}) = \frac{\left(ka\right)}{2}u_{\bm{q}}\chi_n(\bm{q}) \cdot \nonumber\\
&&\cdot\sum_{m \neq n}\left(\bm{\hat{e}_q}\bm{\hat{a}}_{nm}\right)
\left[\exp\left(i\bm{q}\bm{a}_{nm}\right) - 1\right]
\hat{a}_{nm}^{\alpha}\hat{a}_{nm}^{\beta}\;\;\;.\;\;\;\;
\label{eq;unm3plane22}
\end{eqnarray}
Let us, like in the transition from 
(\ref{eq;force02}) to (\ref{eq;force03}), again
assume that we consider crystal lattices with the central symmetry.
For the real part of the stress from (\ref{eq;unm3plane22}) we get:
\begin{eqnarray}
s_{n}^{\alpha\beta}(\bm{q}) = \frac{\left(kau_q\right)}{2}
N_c\Upsilon^{\alpha\beta}_1\left(\bm{q},\bm{\hat{e}}_q\right)
\sin\left(\omega_{\bm{q}} t - \bm{q}\bm{r}_n +\phi_{\bm{q}}\right)\;,\;\;\;\;\;\;\;\;
\label{eq;unm3plane40}
\end{eqnarray}
where
\begin{eqnarray}
\Upsilon^{\alpha\beta}_1\left(\bm{q},\bm{\hat{e}}_q\right)\equiv \frac{1}{N_c}
\sum_{m \neq n}\left(\bm{\hat{e}}_{\bm{q}} \bm{\hat{a}}_{nm}\right)
\sin\left(\bm{q}\bm{a}_{nm}\right)
\hat{a}_{nm}^{\alpha}\hat{a}_{nm}^{\beta}\;\;.\;\;\;\;\;\;\;\;
\label{eq;unm3plane41}
\end{eqnarray}
Formulas (\ref{eq;unm3plane40},\ref{eq;unm3plane41}) express 
local atomic stress elements due to a particular wave
through the parameters of the lattice and the parameters of the propagating wave.

\subsection{Atomic level pressure}

In accord with \cite{Egami19821,Chen19881,Levashov2008B}, 
we define atomic level pressure as:
\begin{eqnarray}
p_n(\bm{q}) = \frac{1}{3v_o}
\left[s^{xx}_n(\bm{q})+s^{yy}_n(\bm{q})+s^{zz}_n(\bm{q})\right]
\;\;\;,\;\;\;\;
\label{eq;pressure01}
\end{eqnarray}
where $v_o$ is atomic volume. Here we will assume that
atomic volume is a constant approximately 
equal to the inverse of the number density, 
i.e., $v_o\approx 1/\rho_o$.

It follows from 
(\ref{eq;unm3plane40},\ref{eq;unm3plane41},\ref{eq;pressure01}) that: 
\begin{eqnarray}
p_{n}(\bm{q}) = \frac{\left(kau_q\right)N_c}{6v_o}
\Upsilon^{p}_1\left(\bm{q},\bm{\hat{e}}_q\right)
\sin\left(\omega_{\bm{q}} t - \bm{q}\bm{r}_n +\phi_{\bm{q}}\right)
\;\;,\;\;\;\;\;\;
\label{eq;pressure02}
\end{eqnarray}
where
\begin{eqnarray}
\Upsilon^{p}_1\left(\bm{q},\bm{\hat{e}}_q\right)=\frac{1}{N_c}
\sum_{m \neq n}\left(\bm{\hat{e}}_{\bm{q}} \bm{\hat{a}}_{nm}\right)
\sin\left(\bm{q}\bm{a}_{nm}\right)\;\;\;.\;\;\;\;
\label{eq;pressure03}
\end{eqnarray}
Summation over $m$ (spherical integration) for a longitudinal wave leads to:
\begin{eqnarray}
\Upsilon^{p}_1\left(L,qa\right)=\left[\frac{\sin(2\xi)-(2\xi)\cos(2\xi)}{(2\xi)^2}\right],\;\;\;
\xi=\frac{qa}{2}\;.\;\;\;\label{eq;pressure04L11}
\end{eqnarray}

It also follows from (\ref{eq;pressure03}) that transverse 
waves do not contribute to the pressure.

\subsection{Mean square of the atomic level pressure} 

It follows from
(\ref{eq;pressure02},\ref{eq;pressure03},\ref{eq;pressure04L}) 
that time averaged square of the pressure due to a longitudinal wave with
the wavevector of magnitude $q$ is: 
\begin{eqnarray}
\left<\left[p_{n}(q)\right]^2\right> = \frac{\left(kau_q\right)^2 N_c^2}{72v_o^2}
\cdot\Upsilon^{p}_2\left(L,qa\right)\;\;\;,\;\;\;\;
\label{eq;pressure05}
\end{eqnarray}
where
\begin{eqnarray}
\Upsilon^{p}_2\left(L,qa\right) \equiv \left[\Upsilon^{p}_1\left(L,qa\right)\right]^2\;\;.\;\;\;\;
\label{eq;pressure04L}
\end{eqnarray}
In the long wavelength approximation:
\begin{eqnarray}
\Upsilon^{p}_{2,lw}\left(L,qa\right)\approx  \frac{1}{9}(qa)^2\;\;\;.\;\;\;
\end{eqnarray}

Expressing $u_q^2$ from 
(\ref{eq;equip3D01}) and then using it
in (\ref{eq;pressure05}) leads, 
after integration (\ref{eq;debye01}) over $q$,
to:
\begin{eqnarray}
\left<p_{n}^2\right> \approx 
k_bT\cdot \left(\frac{ka^2N_c}{36v_o^2}\right)\cdot 0.29\;\;\;\;.
\label{eq;pressure06}
\end{eqnarray}
Calculations in the long wavelength approximation lead to
$\approx 1.11$ instead of $\approx 0.29$.

\subsection{Atomic level pressure energy}

In several previous publications atomic level stress energies
were discussed \cite{Egami19821,Chen19881,Levashov2008B}. 
These quantities are of interest, in particular,
because of their values in the liquid states. 
According to MD simulations, the stress energy for every stress
component is very close to $(1/4)k_bT=(1/6)(3/2)k_b T$. 

It is well known that the average potential energy of a classical $3D$ harmonic 
oscillator is equal to $(3/2)k_bT$. Thus the values of the atomic levels stress energies
are such that it appears that the average potential energy
of {\it some $3D$ harmonic oscillator} is equally divided between 
the six independent components of the atomic level stresses. 
Thus it is interesting to estimate the values of the local atomic stress energies 
in our model.

The expression for the local atomic pressure energy is \cite{Egami19821,Levashov2008B}:
\begin{eqnarray}
<U^p> \equiv \frac{v_o\left<p_{n}^2\right>}{2B} \;\;\;\;.\;\;\;\;\;\;
\label{eq;pressure06}
\end{eqnarray}
In order to evaluate the expression we need to know the value of the bulk
modulus $B$.
The expressions for the elastic constants 
were discussed before \cite{Egami19821,Levashov2008B}.
The results of their evaluations are:
\begin{eqnarray}
B =\frac{\varkappa}{8}\;\;,\;\;\;\;\;\; 
G=\frac{\varkappa}{30}\;\;,\;\;\;\;\;\;
\varkappa = \left(ka^2\right)\frac{N_c}{v_o}\;\;,\;\;\;\;\;\;\; 
\label{eq;elastC01}
\end{eqnarray}
where $G$ is the shear modulus.
Using the value of the bulk modulus $B$
for the average pressure stress energy we get:
\begin{eqnarray}
<U^p> \approx 
\left(\frac{1}{4}\right)k_bT \cdot \left(\frac{1}{7.76}\right)\;\;.
\label{eq;pressure06}
\end{eqnarray}
This energy is significantly smaller than the value of the 
pressure energy that was obtained for liquids.

In the long wavelength approximation we get:
\begin{eqnarray}
<U^p_{lw}> \cong 
\left(\frac{1}{4}\right)k_bT \cdot \frac{1}{2.03}\;\;.
\label{eq;pressure07}
\end{eqnarray}

Thus in the long wavelength approximation the
atomic level pressure stress energy is approximately 2 times 
smaller than the equipartition value, in agreement 
with \cite{Egami19821}.
However, without the long wavelength approximation 
the local atomic pressure energy is more than 7 times smaller than the equipartition
value.
We address these differences further in the discussion
section.

\subsection{Pressure-pressure correlation function}

Our goal here is to address the behavior of 
the atomic level stress correlation function that
is analogous to the function $F(t,r)$ that could be derived 
from the macroscopic Green-Kubo stress correlation function 
and that was studied by MD simulations 
previously \cite{Levashov20111,Levashov2013}.
Thus we introduce: 
\begin{eqnarray}
C^p(t,r)=\left(\frac{a}{2\pi}\right)^3N\int_0^{Q_{max}} C^p(t,r,q)\,4\pi q^2dq\;\;\;,\;\;\;\;
\label{eq;CPcorr01}
\end{eqnarray}
where
\begin{eqnarray}
C^p(t,r,q)=\left<p_{n}(t_o,q)\cdot p_{m}(t_o+t,q)\right>\;\;\;.\;\;\;\;
\label{eq;CPcorr02}
\end{eqnarray}
Spherical averaging and the averaging over $t_o$ are assumed in (\ref{eq;CPcorr02}).

Note that the correlation function that we introduced 
in (\ref{eq;CPcorr01},\ref{eq;CPcorr02}) is the correlation 
function per pair of particles and not the correlation function
between ``a central particle" and ``the particles in the spherical annulus", 
as it was done in \cite{Levashov20111,Levashov2013}. 

\begin{figure}
\begin{center}
\includegraphics[angle=0,width=2.4in]{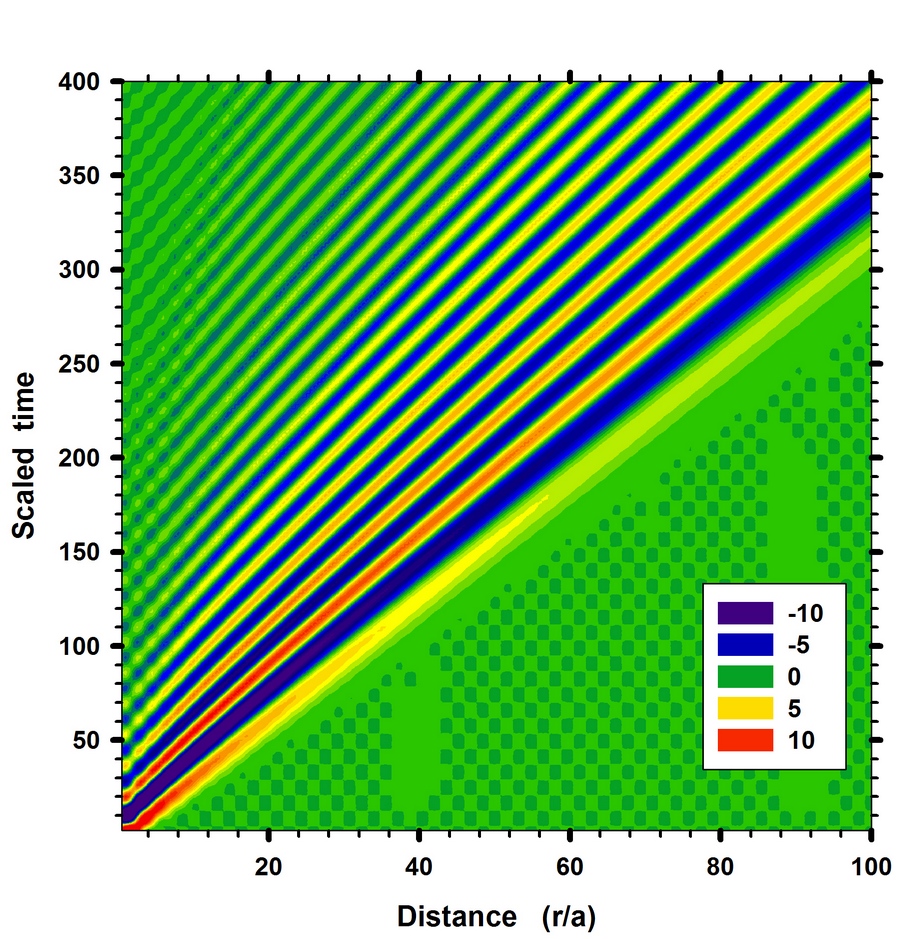}
\caption{Pressure-Pressure correlation function without
long wavelength approximation.
}\label{fig:pp-correlations-1}
\end{center}
\end{figure}

\begin{figure}
\begin{center}
\includegraphics[angle=0,width=2.4in]{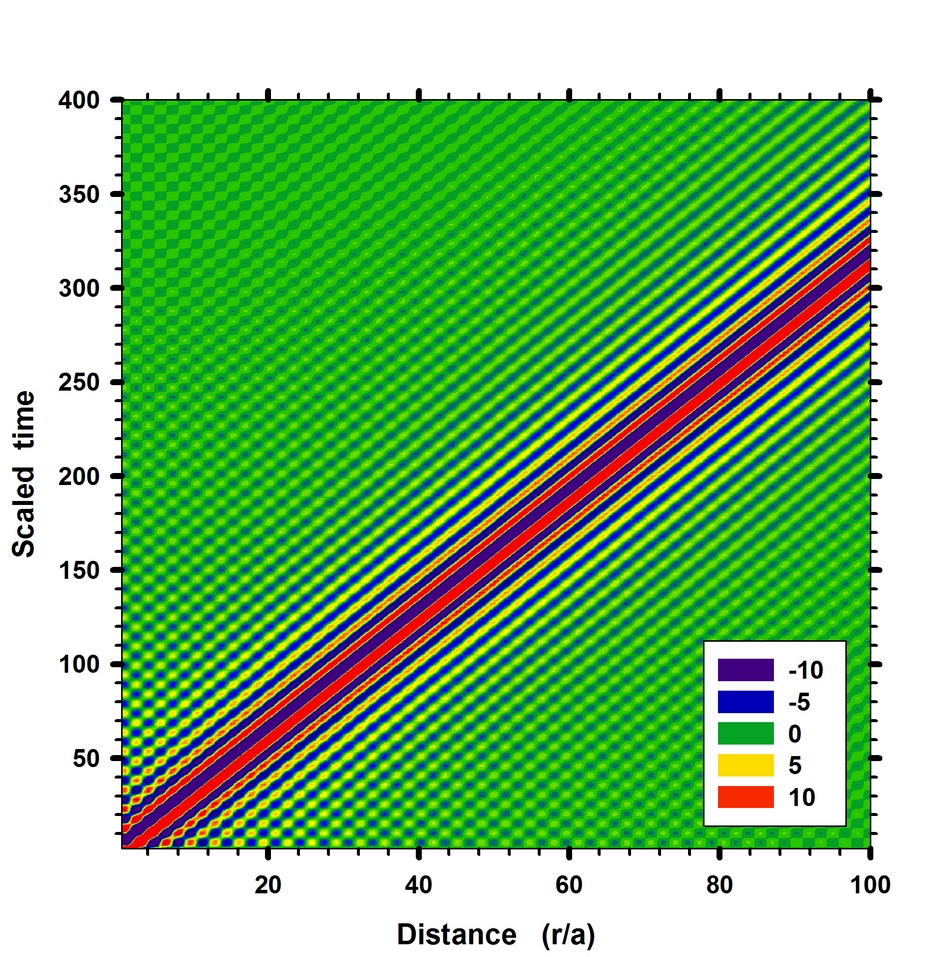}
\caption{Pressure-Pressure correlation function with
long wavelength approximation.
}\label{fig:pp-correlations-lw-1}
\end{center}
\end{figure}

From (\ref{eq;pressure02},\ref{eq;pressure03}) it follows that: 
\begin{eqnarray}
&&C^p(t,r,q)=
\frac{\left(kau_q\right)^2}{36v_o^2}\cdot
\Upsilon^{p}_2\left(L,qa\right)\cdot\;\;\;\;\;\;\;\;\;\\  
&&\left<\sin\left[\omega_{\bm{q}} t_o - \bm{q}\bm{r}_n +\phi_{\bm{q}}\right]
\sin\left[\omega_{\bm{q}} (t_o+t) - \bm{q}\bm{r}_m +\phi_{\bm{q}}\right]\right>
\;\;\;.\;\;\;\;\nonumber
\label{eq;ppcf01}
\end{eqnarray}
From representing the product of {\it sines} as a 
difference of {\it cosines} it follows that one of the {\it cosines} 
gives zero on averaging over $\phi_{\bm{q}}$.
Thus we get:
\begin{eqnarray}
&&C^p(t,r,q)=\nonumber\\
&&\frac{\left(kau_q\right)^2}{36v_o^2}\cdot
\Upsilon^{p}_2\left(L,qa\right)\cdot
\frac{1}{2}\left<\cos\left[\frac{\omega_{\bm{q}} t - \bm{q}\bm{r}_{nm}}{2}\right]\right>
\label{eq;ppcf02}
\end{eqnarray}
Further we rewrite $\cos\left[\frac{\omega_{\bm{q}} t - \bm{q}\bm{r}_{nm}}{2}\right]$ as:
\begin{eqnarray}
\left<
\cos\left[\frac{\omega_{\bm{q}} t}{2}\right]
\cos\left[\frac{\bm{q}\bm{r}_{nm}}{2}\right]
\right>
+
\left<
\sin\left[\frac{\omega_{\bm{q}} t}{2}\right]
\sin\left[\frac{\bm{q}\bm{r}_{nm}}{2}\right]
\right>\;\;\;\;\;\;\;
\label{eq;ppcf03}
\end{eqnarray}
Spherical averaging of the second term over the directions of
$\bm{r}_{nm}$ is zero.
Spherical averaging in the first term gives:
\begin{eqnarray}
\left<\cos\left[\frac{\bm{q}\bm{r}_{nm}}{2}\right]\right>=
2\frac{\sin\left(qr/2\right)}{(qr/2)}\;\;\;.
\label{eq;ppcf04}
\end{eqnarray}
Using the expression (\ref{eq;equip3D01}) for $u_q^2$ 
we rewrite (\ref{eq;ppcf02}) as:
\begin{eqnarray}
&&C^p(t,r,q)=k_bT\left(\frac{ka^2N_c}{v_o^2}\right)\cdot\frac{1}{36}\cdot\nonumber\\
&&\cdot\frac{2}{N}\left\{\frac{\Upsilon^{p}_2\left(L,qa\right)}{D_L(qa)}\right\}\cdot
\left\{\frac{\cos\left(\omega_{q} t/2\right)\sin\left(qr/2\right)}{(qr/2)}\right\}\;\;,\;\;\;\;\;
\label{eq;ppcf05}
\end{eqnarray}
where, according to (\ref{eq;dispersion3D03}), $\omega_q = \omega_o \sqrt{D_L(qa)}$.
The product of the {\it cosine} and {\it sine} in (\ref{eq;ppcf05}) 
could be rewritten as:
\begin{eqnarray}
\frac{1}{2}\left[\sin\left(\frac{qr-\omega_{q} t}{2}\right)+\sin\left(\frac{qr+\omega_{q} t}{2}\right)\right]\;\;\;.\;\;\;\;\;\;
\label{eq;ppcf06}
\end{eqnarray}
The first {\it sine} corresponds to a wave propagating from the central particle.
This {\it sine} is zero when $r-(\omega_{\bm{q}}/q)t=0$.
The argument of the second {\it sine} is always positive for positive
$t$ and $r$. 
For positive times the contribution to the stress correlation function due to all waves 
from the second {\it sine} is much smaller than from the first {\it sine}. 
However, for negative times the second {\it sine} behaves like the first 
{\it sine} for positive times. 

In order to find the pressure correlation function due to all waves
we have to integrate (\ref{eq;ppcf05}) over all $q$ using (\ref{eq;debye01}).
Fig.\ref{fig:pp-correlations-1} shows
the results for the pressure-pressure correlation function due to all
waves without long wavelength approximation. 
Fig.\ref{fig:pp-correlations-lw-1} 
shows the result with long wavelength approximation.

\subsection{An example of the local atomic shear stress and shear stress energy}

In accord with references \cite{Egami19821,Chen19881,Levashov2008B} we define:
\begin{eqnarray}
\sigma^{\epsilon}_n(\bm{q},\bm{\hat{e}}_{\bm{q}}) = 
\left(\frac{\sqrt{2}}{v_o}\right)s^{xy}_n(\bm{q},\bm{\hat{e}}_{\bm{q}})\;\;\;.\;\;\;\;\;
\label{eq;Sepsilon01}
\end{eqnarray}
Both longitudinal and transverse waves contribute to 
$\sigma^{\epsilon}_n(\bm{q},\bm{\hat{e}}_{\bm{q}})$. 
Their contributions depend on the magnitude and direction of $\bm{q}$ and the direction
of $\bm{\hat{e}}_{\bm{q}}$.
Below, for shortness, we present the formulas for the transverse waves only.
The formulas for the longitudinal waves are analogous.

From 
(\ref{eq;unm3plane40},\ref{eq;unm3plane41},\ref{eq;Sepsilon01},\ref{eq;equip3D01}) we get:
\begin{eqnarray}
&&\left<\left[\sigma^{\epsilon}_n(T,q)\right]^2\right> = 
\left(k_bT\right)\left(\frac{ka^2 N_c}{2v_o^2}\right)\frac{1}{N}
\left[\frac{\Upsilon^{xy}_2\left(T,qa\right)}{D_T\left(qa\right)}\right],\;\;\;\;\;
\label{eq;Sepsilon02}
\end{eqnarray}
where:
\begin{eqnarray}
\Upsilon^{xy}_2\left(T,qa\right) \equiv 
\left<\left[\Upsilon^{xy}_1\left(T,\bm{q},\bm{\hat{e}_q},a\right)\right]^2\right> 
\label{eq;Sepsilon03}\;,
\end{eqnarray}
and $\Upsilon^{xy}_1\left(T,\bm{q},\bm{\hat{e}_q},a\right)$ is given by 
(\ref{eq;unm3plane41}).
The averaging in (\ref{eq;Sepsilon03}) is over all directions
of $\bm{\hat{e}}_{\bm{q}}$ orthogonal to $\bm{q}$ 
and then over the directions of $\bm{q}$.

\begin{figure}
\begin{center}
\includegraphics[angle=0,width=3.3in]{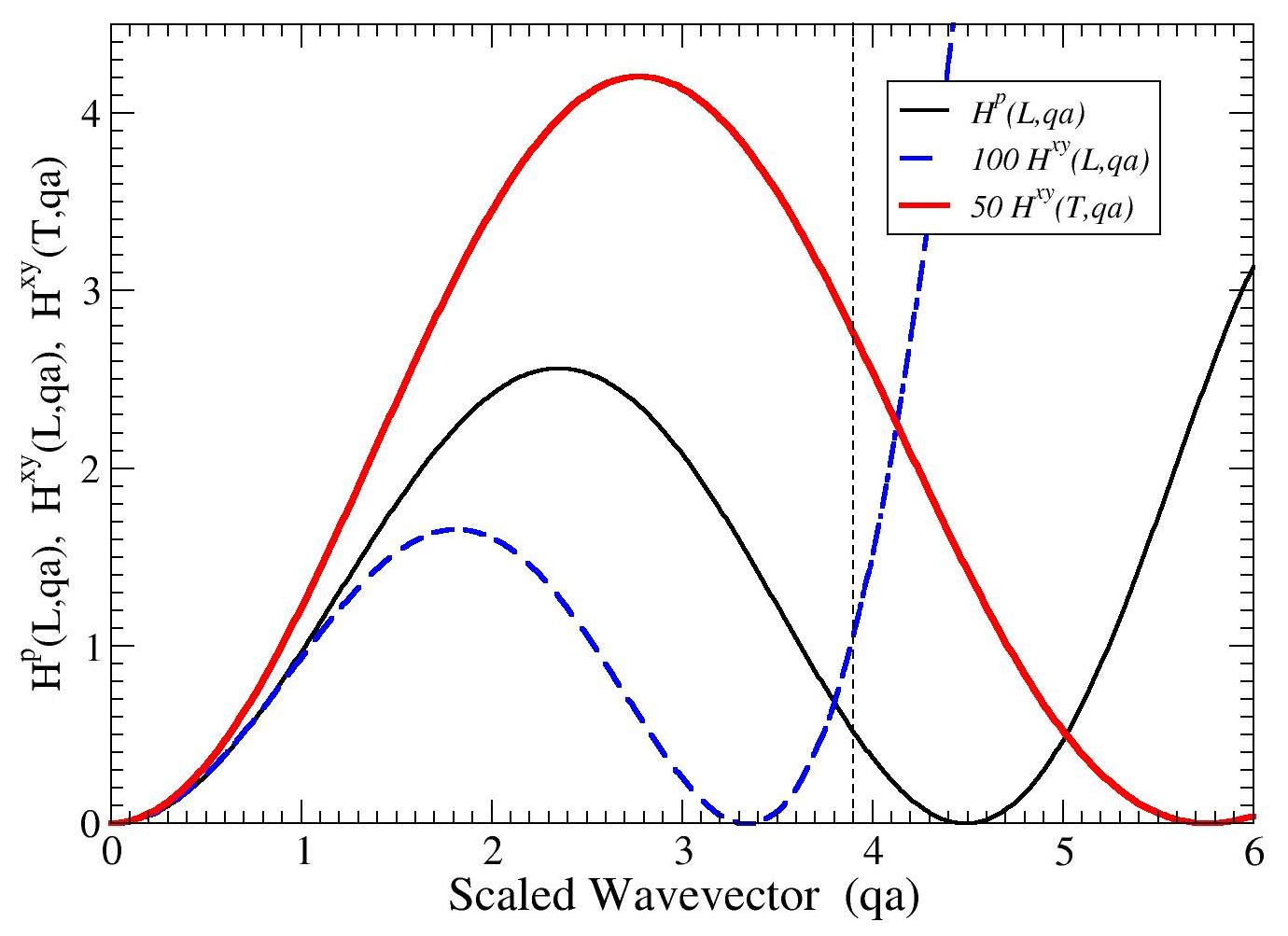}
\caption{Functions $H^p(L,qa)$, 
$H^{xy}(L,qa)$, and $H^{xy}(T,qa)$
from (\ref{eq;Hqa031},\ref{eq;Hqa032},\ref{eq;Hqa033}).
Note that $H^{xy}(L,qa)$ curve 
was scaled by $100$, 
while $H^{xy}(T,qa)$ curve was scaled by
$50$.
Thus the contribution of the transverse waves to the average 
square of the shear stress is significantly larger than the 
contribution from the longitudinal waves. Also note that there are
two polarizations of the transverse waves, 
while the figure shows the contribution from one polarization only.
}\label{fig:with-q2-1-crop}
\end{center}
\end{figure}

We were not able to produce analytical expressions for 
$\Upsilon^{xy}_2\left(T,qa\right)$ and $\Upsilon^{xy}_2\left(L,qa\right)$. 
However, we calculated them numerically \cite{numerically}.
Fig.\ref{fig:with-q2-1-crop} shows the dependencies of
\begin{eqnarray} 
H^p(L,qa)\equiv&&\left[\frac{\Upsilon^{p}_{2}(L,qa)}{D_L(qa)}\right](qa)^2\;\;\;,\;\;\;\;
\label{eq;Hqa031}\\ 
H^{xy}(L,qa)\equiv&&\left[\frac{\Upsilon^{xy}_{2}(L,qa)}{D_L(qa)}\right](qa)^2\;\;\;,\;\;\;\;
\label{eq;Hqa032}\\
H^{xy}(T,qa)\equiv&&\left[\frac{\Upsilon^{xy}_{2}(T,qa)}{D_T(qa)}\right](qa)^2\;
\label{eq;Hqa033}
\end{eqnarray}
on $qa$ without long wavelength approximation.
In (\ref{eq;Hqa031},\ref{eq;Hqa032},\ref{eq;Hqa033}) 
we introduced the factor $(qa)^2$ 
assuming further integrations over $q$. 

We also obtained analytical expressions for 
(\ref{eq;Sepsilon03}) in the long wavelength approximation:
\begin{eqnarray}
\frac{\Upsilon^{xy}_{2,lw}(L,qa)}{D_L(qa)} \approx \left[\frac{8}{675}\right],\;\;\;\;\;
\frac{\Upsilon^{xy}_{2,lw}(T,qa)}{D_T(qa)} \approx \left[\frac{18}{675}\right].\;\;\;\;\;
\label{eq;epsilon03}
\end{eqnarray}

In order to evaluate mean square stresses due to all waves we 
have to integrate (\ref{eq;Sepsilon02}), i.e., the curves in
Fig.\ref{fig:with-q2-1-crop}, over all $q$ using (\ref{eq;debye01}).
The results of these integrations, expressed
in terms of the average energy of the atomic level shear stresses, are: 
\begin{eqnarray}
&&\frac{v_o\left<\left[\sigma_{n}^{\epsilon}(L,T)\right]^2\right>}{4G} \approx 
\left(\frac{1}{4}\right)k_bT \cdot \tau(L,T)\;\;\;,\;\;\;\label{eq;epsilon041}\\
&&\tau(L) \approx \frac{3.1}{135}, \;\;\;\;\;\tau(T)\approx\frac{20.6}{135}\;\;\;.\;\;\;
\nonumber
\end{eqnarray}
The shear stress energy coefficient due to all waves is:
\begin{eqnarray}
\tau(L) + 2\tau(T) \approx (44.3/135) \approx \frac{1}{3}\;\;\;.\;\;
\label{eq;epsilon042}
\end{eqnarray}
Thus we got the result which is 3 times smaller than the equipartition
result for certain MD liquids \cite{Chen19881,Levashov2008B}.
Also note that this result is more than two times larger than the 
result that was obtained for the pressure (\ref{eq;pressure06}).

In the long wavelength approximation we get:
\begin{eqnarray}
\tau_{lw}(L) = \frac{24}{135}, \;\;\;\;\;\;\;\;\;\tau_{lw}(T)=\frac{54}{135},\;\;\;\\
\tau_{lw}(L) + 2\tau_{lw}(T) = (132/135) \approx 1\;.\;\;\;
\label{eq;epsilon043}
\end{eqnarray}
Thus in the long wavelength approximation
we essentially have $(1/4)k_bT$ dependence.

\subsection{Shear Stress Correlation Function}

In order to introduce a shear stress correlation function which is analogous to the 
function $F(t,r)$ in \cite{Levashov20111,Levashov2013} we first 
introduce a correlation function due to a particular 
wave. For a particular transverse wave we write:
\begin{eqnarray}
&&C^{\epsilon}_{\bm{q}}(T,t,r,\bm{q},\bm{\hat{e}_q})=
\left(\frac{2}{v_o^2}\right)\cdot\\
&&\left<s_{n}^{xy}(t_o,\bm{q},\bm{\hat{e}_q})
\cdot s_{m}^{xy}(t_o+t,\bm{q},\bm{\hat{e}_q})\right>=\\
&&\left(k_bT\right)\left(\frac{ka^2 N_c}{2v_o^2}\right)\frac{1}{N}
\left[\frac{\left(\Upsilon^{xy}_1\left(T,\bm{q},\bm{\hat{e}_q},a\right)\right)^2}{D_T\left(qa\right)}\right]
\cdot\\
&&\left<\sin\left[\omega_{\bm{q}} t_o - \bm{q}\bm{r}_n +\phi_{\bm{q}}\right]
\sin\left[\omega_{\bm{q}} (t_o+t) - \bm{q}\bm{r}_m +\phi_{\bm{q}}\right]\right>\;.
\;\;\;\;\;\;\;\;
\label{eq;CXYcorr02}
\end{eqnarray}
Similarly to how it was done in the transition from 
(\ref{eq;ppcf01}) to (\ref{eq;ppcf05}) for the pressure-pressure
correlation function we now average over the 
different directions of $\bm{r}_{nm}$.
The difference with the pressure-pressure case is that
now the prefactor depends on the direction and the polarization of the wave, 
while in the pressure-pressure case it depends only on the magnitude of
the wavevector. This difference is, however, irrelevant for the averaging over 
the directions of $\bm{r}_{nm}$. Then we perform the averaging
over the polarization and the direction of the wave. 
This averaging is identical to the averaging that was done in derivations 
of (\ref{eq;Sepsilon02},\ref{eq;Sepsilon03}). 
Thus we get:
\begin{eqnarray}
C^{\epsilon}(T,t,r)=\left(\frac{a}{2\pi}\right)^3N\int_0^{Q_{max}} C^{\epsilon}(T,t,r,q)\,4\pi q^2dq
\;,\;\;\;\;\;\;\;
\label{eq;CXYcorr03}
\end{eqnarray}
where
\begin{eqnarray}
&&C^{\epsilon}(T,t,r,q)=k_bT\left(\frac{ka^2N_c}{v_o^2}\right)\cdot E(\omega_q t)\cdot\nonumber\\
&&\cdot\frac{2}{N}\left\{\frac{\Upsilon^{xy}_2\left(T,qa\right)}{D_T(qa)}\right\}\cdot
\left\{\frac{\cos\left(\omega_{q} t/2\right)\sin\left(qr/2\right)}{(qr/2)}\right\}\;\;\;.\;\;\;\;
\label{eq;CXYcorr04}
\end{eqnarray}
In (\ref{eq;CXYcorr04}) we introduced the function $E(\omega_q t)$ artificially.
It should not be there if the waves do not decay with time or distance.
However, if we want to make a comparison with the stress correlation
functions calculated in MD simulations on liquids, then it is reasonable to
assume that waves decay. For the sake of a qualitative comparison we
assume that \cite{Mizuno2013}:
\begin{eqnarray}
E(\omega_q, t) = \exp\left[-0.3(\omega_q/\omega_o)^2 \omega_o t\right]\;\;\;.\;\;\;
\label{eq;efun}
\end{eqnarray}

\begin{figure}
\begin{center}
\includegraphics[angle=0,width=3.0in]{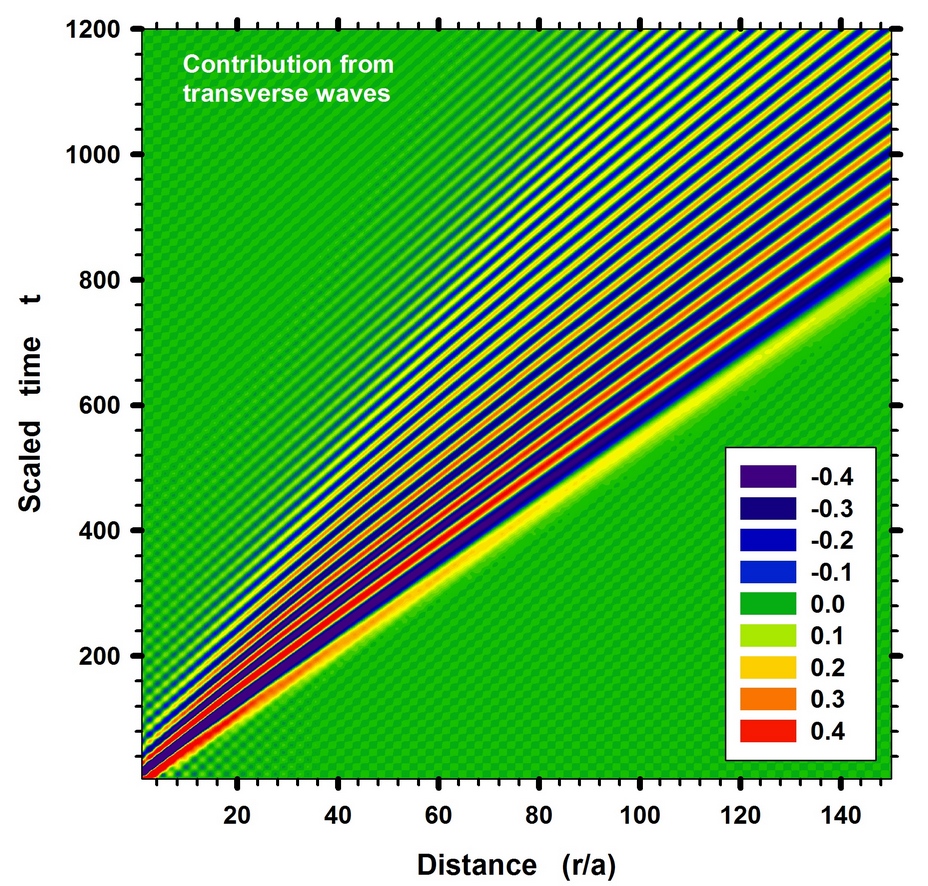}
\caption{Contribution from one polarization of the transverse waves to the shear stress
correlation function. Scaled time is $\omega_o t$. 
No damping of the waves is assumed.}
\label{fig:sscf-xy-T-total-1-1}
\end{center}
\end{figure}

Figure \ref{fig:sscf-xy-T-total-1-1} shows the stress correlation function calculated
numerically from (\ref{eq;CXYcorr03},\ref{eq;CXYcorr04}) under the assumption that the first row in
(\ref{eq;CXYcorr04}) is equal to 1. 
Panel (a) of Figure \ref{fig:sscf-joined-1} again shows the stress correlation function
(\ref{eq;CXYcorr03},\ref{eq;CXYcorr04}), but now with  $E(\omega_q, t)$ given by (\ref{eq;efun}).
We do not show in the figure the contribution from the longitudinal waves.
This contribution is qualitatively similar to the contribution from the transverse waves.
However, this contribution is significantly smaller in magnitude.
Also, since the speeds of the longitudinal waves are $\approx \sqrt{3}$ times larger than 
the speeds of the transverse waves,
the diagonal lines in the contribution from the longitudinal waves 
have slopes which are $\approx \sqrt{3}$ times smaller than the slopes
of the diagonal lines from the transverse waves.

It is interesting to compare the panel (a) of Fig.\ref{fig:sscf-joined-1}
with the panel (b) of Fig.4 in Ref.\cite{Levashov2013}.
Note that in the present paper we changed the axes and now 
$x$-axis shows the distance, while $y$-axis shows the time.
In the panel (b) of Fig.4 in Ref. \cite{Levashov2013} we see
two waves. One wave is longitudinal and another wave is transverse.
There we also see \pdf - like contribution to the stress correlation function.
In the panel (a) of Fig.\ref{fig:sscf-joined-1} we see the contribution 
from the transverse waves only,
since we did not include in it the contribution from the longitudinal waves or the
\pdf - like structure. 
Besides the differences mentioned above, it is clear that the contribution
to the stress correlation function from the transverse waves
observed in MD simulation and in the present calculations are qualitatively similar.
In particular, if we consider how intensity changes with increase 
of time for a given distance we observe 
at first positive intensity and then negative intensity.

Formula (\ref{eq;ppcf06}) suggests that the speed of the wave
corresponds to the slope of the first boundary between the positive and negative 
intensities. This interpretation is different from the one adopted in 
Ref.\cite{Levashov2013}. There it was assumed that the center of the wave
corresponds to the maximum of the positive intensity. 
With the new interpretation the speed of the longitudinal waves
in the panel (b) of Fig.4 in Ref. \cite{Levashov2013} is $c_l \approx 7500$ (m/s) 
while before it was argued that it is 6000 (m/s). The new speed of the
transverse waves is $c_t \approx 5000$ (m/s), 
while before it was argued that it is 3000 (m/s).
Note that, according to the new values, $c_l/c_t \approx 1.5$, 
which is not quite $\sqrt{3}$.

\subsection{Fourier transforms of the shear stress correlation function}

The atomic level stress correlation functions, 
like those in (\ref{eq;ppcf05},\ref{eq;CXYcorr04}), 
could be calculated in MD simulations \cite{Levashov20111,Levashov2013}. 
In this section we analyze what information 
could be obtained by performing Fourier transforms of 
these stress correlation functions (\ref{eq;ppcf05},\ref{eq;CXYcorr04}). 
Thus further we consider a function which is structurally similar to the 
stress correlation functions (\ref{eq;ppcf05},\ref{eq;CXYcorr04}):
\begin{eqnarray}
f(t,r) \equiv \int_0^{Q_{max}} h(q,t)
\cos\left(\frac{\omega_q t}{2}\right)
\sin\left(\frac{qr}{2}\right)dq\;\;\;.\;\;\;\;\;\;
\label{eq;ftr11}
\end{eqnarray}
For the shear stress correlation function due 
to the transverse waves (\ref{eq;CXYcorr04}), for example, we have:
\begin{eqnarray}
&&f(t,r) \equiv r\cdot C^{\epsilon}(T,t,r)\label{eq;ftr121}\;\;\;,\;\;\;\\
&&h(q,t)\equiv
\alpha \left(\frac{Y_2^{xy}(T,q)}{D_T(qa)}\right)\cdot E(\omega_q, t)\cdot q\;\;\;,\;\;\;
\label{eq;ftr12}
\end{eqnarray}
where $\alpha$ is a numerical coefficient. Note that $f(t,r)$, as we define it, is the correlation function per
pair of particles multiplied by $r$.

Further we define:
\begin{eqnarray}
&&\tilde{f}(t,q) \equiv \int_0^{\infty} f(t,r)\,\sin(qr)dr \;\;\;,\;\;\;\label{eq;ftr21x}\\
&&\tilde{f}(\omega,r) \equiv \int_0^{\infty} f(t,r)\,\cos(\omega\,t)dt \;\;\;,\;\;\;\label{eq;ftr22x}\\
&&\tilde{f}(\omega,q) \equiv \int_0^{\infty}\int_0^{\infty} f(t,r)\,\cos(\omega\,t)\,\sin(qr)\,dt\,dr \;\;\;.\;\;\;\;\;\label{eq;ftr23x}
\end{eqnarray}
From (\ref{eq;ftr21x},\ref{eq;ftr11},\ref{eq;ftr12}) we get:
\begin{eqnarray}
\tilde{f}(t,q) \equiv \left(\frac{\pi}{2}\right) h(2q,t)
\cos\left(\frac{\omega_{2q}t}{2}\right)\;\;\;\;,\label{eq;ftr21}\;\;\;\;
\end{eqnarray}
where $\omega_{2q}^2=\omega_o^2 D_T(2qa)$.
Thus $\tilde{f}(t,q) $, for every value of $q$, oscillates in time with the period 
determined by the dispersion relation.
The decrease in the amplitude of oscillations with increase of time
is determined by the damping function $E(\omega_q, t)$ 
(\ref{eq;ftr12}). If there were no damping function the amplitude of oscillations
would remain constant. 

The situation with the Fourier transform of $f(t,r)$ over time is more complicated.
In general, for liquids the damping function is not exponential, but a
function that describes different relaxation regimes. For an exponential damping
function it is possible to perform the Fourier transform of the integrand 
in (\ref{eq;ftr11})  in a closed analytical form. 
However, then it is still necessary to integrate the obtained analytical expression over $q$.
Here we will not consider the case with damping in more detail.
If there is no damping, i.e., if $E(\omega_q, t)=1$, then
from (\ref{eq;ftr22x},\ref{eq;ftr11},\ref{eq;ftr12}) 
the Fourier transform of $f(t,r)$ over $t$ is:
\begin{eqnarray}
\tilde{f}(\omega,r) \equiv \left(\frac{\pi}{2}\right) h(q_{2\omega})
\sin\left(\frac{q_{2\omega}r}{2}\right)\;\;\;,\;\;\;\label{eq;ftr22}
\end{eqnarray}
where $(2\omega)^2=\omega_o^2D_T(q_{2\omega}a)$.
Thus, in the absence of damping, the Fourier transform of $f(t,r)$ over 
time (\ref{eq;ftr22}) 
should exhibit for every frequency constant amplitude oscillations 
with wavelength determined by the dispersion relation. 

In the absence of damping, the Fourier transforms of $f(t,r)$ over
$r$ and $t$  (\ref{eq;ftr23x}) lead to:
\begin{eqnarray}
\tilde{f}(\omega,q) \equiv \left(\frac{\pi}{2}\right)^2 h(2q)
\delta\left(\omega-\frac{\omega_{2q}}{2}\right)\;\;,\label{eq;ftr23}\;\;\;
\end{eqnarray}
i.e., to the dispersion relation.
It is clear from  (\ref{eq;ftr21}) that damping should lead to the broadening
of the $\delta$-function in (\ref{eq;ftr23}) .

\begin{figure*}
\begin{center}
\includegraphics[angle=0,width=6.0in]{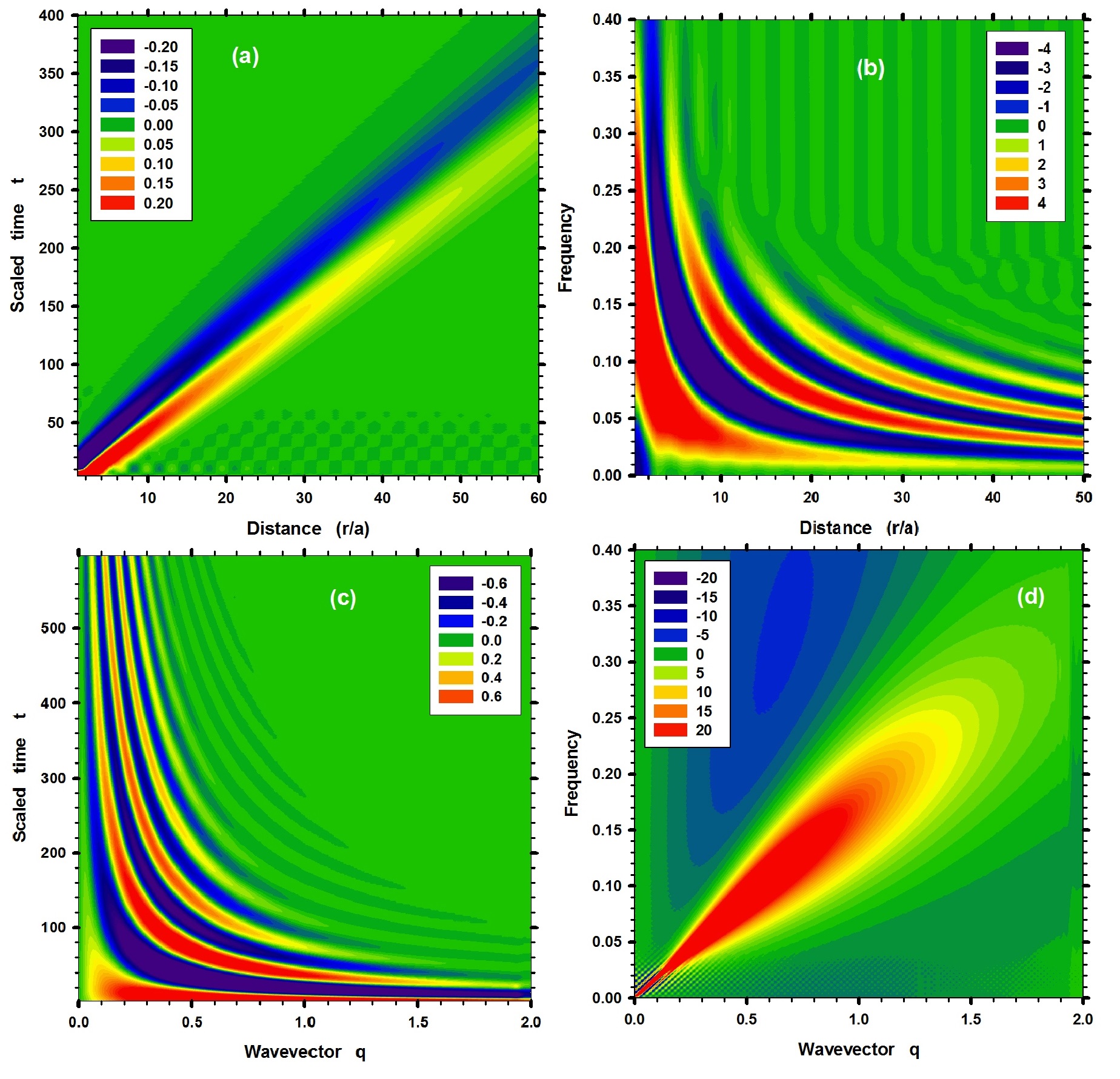}
\caption{Scaled shear stress correlation function, $f(t,r)$ 
(\ref{eq;ftr121}), due to the transverse waves
with exponential damping in time and its Fourier transforms. 
(a) $r$-scaled shear stress correlation function
(\ref{eq;CXYcorr04},\ref{eq;ftr121}) integrated
over all $q$ with $E(\omega_q, t)$ 
given by (\ref{eq;efun}). It was assumed that the numerical prefactor is equal to 1. 
Scaled time is $\omega_o t$.
(b) Time to frequency Fourier transform of the scaled stress correlation function
in the panel (a). Angular frequency is measured in units of $\omega_o$.
(c) Distance to wavevector Fourier transform of the scaled stress correlation function
in the panel (a). The unit of the wavevector is $1/a$.
(d) Time to frequency and distance to wavevector Fourier transforms of the scaled stress
correlation function in the panel (a). 
Note that the center of the diagonal high intensity region
follows the dispersion curve for the transverse waves in Fig.\ref{fig:dispersion02}.
It is clear that the broadening of the dispersion curve is caused 
by the exponential damping (\ref{eq;efun}).}
\label{fig:sscf-joined-1}
\end{center}
\end{figure*}

Panel (a) of Fig.\ref{fig:sscf-joined-1} shows 
the function $f(t,r)$ from
(\ref{eq;ftr121}). Shear stress correlation function,
$C^{\epsilon}(T,t,r)$,  was obtained from
(\ref{eq;CXYcorr04}) by integration over all $q$ with $E(\omega_q, t)$ 
given by (\ref{eq;efun}). It was assumed that $\alpha = 1$.
Panels (c,b,d) show the Fourier transforms 
(\ref{eq;ftr21x},\ref{eq;ftr22x},\ref{eq;ftr23x}) 
of the function $f(t,r)$.

\subsection{Transverse current correlation function}

In agreement with the previous definitions 
\cite{HansenJP20061,EvansDJ19901,Boon19911,Tanaka20091,Tanaka20081,Mizuno2013,Mountain19821}, 
we are using the 
following expression for the real
part of the transverse current:
\begin{eqnarray}
J_T(\bm{k},t) = \sum_i \bm{v}^T_i(t)\cdot 
\cos\left[\bm{k}\bm{r}_i(t)\right]\;\;,
\label{eq;jt01}
\end{eqnarray}
where $\bm{v}^T_i(t) = \bm{v}_i(t) - (\bm{v}_i(t)\bm{\hat{k}})$.
For the contribution from
a transverse wave with the wavevector $\bm{q}$ and the polarization $\bm{\hat{e}_q}$ from 
(\ref{eq;jt01},\ref{eq;poten3dx21},\ref{eq;poten3dx22}) we get:
\begin{eqnarray}
&&J_T(\bm{k},\bm{q},\bm{\hat{e}_q},t) = - u_q \bm{\hat{e}}_q\omega_q B_1(\bm{k},\bm{q},t)\;,\;\nonumber\\
&&B_1(\bm{k},\bm{q})\equiv\sum_i
\sin\left[\omega_q t -\bm{q}\bm{r}_i(t) +\phi_q\right]
\cos\left[\bm{k}\bm{r}_i(t)\right]\;.\;\;\;\;
\label{eq;jt02}
\end{eqnarray}
Then for the correlation 
function due to this wave we have:
\begin{eqnarray}
C_{JT}(\bm{k},\bm{q},\bm{\hat{e}_q},t) \equiv \frac{1}{N}
\left< J_T(...,t_o)J_T(...,t_o+t)\right>\;.\;\;\;
\label{eq;tccf01}
\end{eqnarray}
The averaging in (\ref{eq;tccf01}) is over the initial time $t_o$. 

Using the same logic that was used in the derivations 
of (\ref{eq;ppcf05},\ref{eq;CXYcorr04}) from 
(\ref{eq;jt02},\ref{eq;tccf01},\ref{eq;equip3D01}) we get:
\begin{eqnarray}
&&C_{JT}(\bm{k},\bm{q},\bm{\hat{e}_q},t) = 
\left(\frac{C_e}{N}\right)
\cos\left(\omega_q t\right)\left[X_1 + X_2\right]
\;,\;\;\label{eq;tccf011}\\
&&C_e \equiv \frac{u_q^2 \omega_q^2}{4}=\frac{k_bT}{2NM}\;\;,\;\;\;
X_{1,2} \equiv \sum_{ij}\frac{\sin\left(|\bm{k}\pm \bm{q}|r_{ij}\right)}
{|\bm{k}-\bm{q}|r_{ij}}\;.\;\;\;\;\;\;
\label{eq;tccf012}
\end{eqnarray}
In derivations of (\ref{eq;tccf011},\ref{eq;tccf012}) there also appear two 
other terms which, however, vanish in the limit $N\rightarrow \infty$.
After the integration over $\bm{q}$ contributions from $X_1$ and $X_2$ 
terms are equal to each other.
Note that if $\bm{k}=0$ then the structure of (\ref{eq;tccf011},\ref{eq;tccf012}) 
is rather similar to the structures
of (\ref{eq;ppcf05},\ref{eq;CXYcorr04}).

It follows from (\ref{eq;tccf011},\ref{eq;tccf012}) that it is possible to
introduce and consider {\it atomic level} transverse current correlation function
similarly to how it was done for the atomic level stress correlation function
in Ref. \cite{Levashov20111,Levashov2013}.

\section{Discussion \label{sec:discussion}}

The primary goal of this paper is to gain an insight into the connection
between the atomic level vibrational dynamics
and the atomic level Green-Kubo stress correlation function.
The understanding of this connection is needed to interpret the results
of the previous MD simulations of a 
model liquid \cite{Levashov20111,Levashov2013}.
For this purpose, we considered a simple model in which vibrations are plane waves.
Such representation of vibrations does not imply that we think that vibrations
in liquids or glasses are plane waves. 
The situation in disordered materials is much more complex  
\cite{KeyesT19971,Taraskin2000,Taraskin2002,Scopigno20071}. 
However, the model that we consider is solvable, and it provides 
the needed insight and a recipe for the analysis in Fourier space 
of the atomic level stress correlation functions 
obtained in MD simulations \cite{Levashov20141}. 

Atoms, as they move, do not decompose their motions 
into the orthogonal vibrational modes.
Instead, they experience forces and stresses.
From this perspective, the comparisons of the atomic 
level stress correlation functions 
from different liquids and temperatures 
may provide valuable and, probably, more physical insights into 
the atomic scale dynamics than 
the considerations of the vibrational eigenmodes. 

Further we discuss why atomic level stress energies obtained 
in MD simulations are significantly larger than the values obtained in this paper. 
In the framework of the studied model
the average atomic level stress energies in the long wavelength 
approximation and at high temperatures are similar to those 
in Ref.\cite{Egami19821}. 

To summarize, if the long wavelength approximation is not assumed then
the pressure/shear stress energy within the model is approximately 8/3 times
smaller than the value obtained from MD simulations \cite{Chen19881,Levashov2008B}.
If the long wavelength approximation is assumed then 
the pressure stress energy is 2 times smaller than the value from
MD simulations, while the shear stress energy is approximately equal to
the value from MD. 
Please, see also the text preceding to equation (\ref{eq;elastC01}).

The dependencies of the atomic level stress energies on temperature, 
obtained in MD simulations on a liquid and its glass, can be found in Fig.3
of Ref.\cite{Chen19881} and in Fig.5 of Ref.\cite{Levashov2008B}.
If $T \gtrsim 1000$ (K) the system is in a liquid state. If $T \lesssim 1000$ (K),
the system is in a glass state. 

Note, in the figures, that at $T=0$ (K) 
the atomic level stress energies have a finite value $U_o$.
At $T=0$ (K) there are no vibrations in the classical systems.
Thus the values of the atomic level stress energies  at $T=0$ (K)
are determined only by the structural disorder, 
which also includes variations in the coordination numbers 
between the different atoms, 
as can be seen in Fig.2 of Ref.\cite{Levashov2008E}.

In our present calculations of the atomic level stress energies, 
it was assumed that every
atom interacts with $N_c$ neighbors. 
It was also assumed that $N_c$ is the same for every
atom. But in the previous calculations of the atomic level stress energies with MD simulations,
no distinction was made between the atoms with different 
coordination numbers \cite{Chen19881,Levashov2008B}. 

Formulas (\ref{eq;pressure06},\ref{eq;Sepsilon02}) show that 
average squares of the atomic level stresses are proportional to $N_c$,
while (\ref{eq;elastC01}) shows that elastic constants are also proportional
to $N_c$. 
From this perspective, if all atoms have the same coordination,
the atomic level stress energies, which are proportional to the ratio
of the average squares of the stresses to the relevant elastic constants,
should not exhibit dependence on $N_c$. 
However, in the previous MD simulations, the average squares of 
the stresses and the average values
of the elastic constants were obtained by averaging over the atoms with
different $N_c$.
Thus a (large) part of the average atomic level stress energies obtained in previous
MD simulations in the glass and in the liquid states
might be related to the variations in the coordination numbers between the different atoms. 
Another possibility might be associated with inapplicability of the spherical approximation
for certain coordination numbers or with inapplicability of the plane wave approximation.

Atomic level stresses were originally applied to a model of metallic glass in 
order to describe structural disorder \cite{Egami1980}. 
Later it was briefly discussed that it might be possible to speak about 
structural and vibrational contributions to the atomic level stresses \cite{Egami19821}. 
However, no systematic attempts to separate structural and vibrational 
contributions to the atomic level stresses were previously made. 
Considerations of the atomic level stress energies for the subsets of atoms with different 
coordination numbers should help to elucidate the roles of structural disorder 
and vibrational dynamics. 
This is in agreement with the results of \cite{Iwashita2013}.

In the glass state, dependencies of the atomic level stress energies on temperature
(obtained in MD simulations) can approximately be described by a formula: $U_o + (1/6)k_b T$. 
It is natural to associate $U_o$ with the structural contribution, while $(1/6)k_b T$ 
with the vibrational contribution. 
Note that the rates of increase of the atomic level stress energies in the glass 
obtained from MD simulations are larger than the values that 
we derived in this paper 
(see formulas (\ref{eq;pressure06}) and (\ref{eq;epsilon041})).
Note also that, according to the present paper, the rates of increase of the
pressure and shear energies should be different.
However, previous MD simulations show similar rates.
This can reflect the fact that vibrations in disordered media are not plane waves
as we assumed here. It also can reflect the fact that 
structural disorder changes even in the glass state.
For example, as temperature increases (still in the glass state) 
there is a weak change in the number of atoms with a given
coordination  as can be seen in Fig.2 of Ref.\cite{Levashov2008E}.
However, this weak change can also be a consequence of the vibrational dynamics.

In the liquid state temperature dependence of the atomic level stress energies
closely follows $(1/4)k_b T$. Absence of $U_o$ in the last formula suggests that
in the liquid state there is no ``frozen in" structural contribution. 
Thus Fig.3 of Ref.\cite{Chen19881} and Fig.5 of Ref.\cite{Levashov2008B}
suggest that it might be possible to speak about vibrational and configurational
degrees of freedom in the glass state. 
However, in the liquid states these degrees of freedom
appear to be completely mixed.

In derivations of the $(1/4)k_b T$ law for the atomic level stress energies,
it was assumed that the values of atomic level stresses can vary from $-\infty$ to $+\infty$ 
\cite{Egami19821,Levashov2008B}. 
Any, in magnitude, contribution to the atomic level stress can come from the repulsive
part of the potential for a given coordination number because very large forces can originate from 
the repulsive core. 
On the other hand, only a limited contribution
can come from the attractive part of the potential for a given coordination number.
This also can be the reason why the equipartition result does not hold for a fixed
coordination number, but holds when averaging is done over all coordination numbers.

Derivation of the equipartition from the Boltzmann distribution 
\cite{Egami19821,Levashov2008B} implies the presence
of ergodicity for every central atom and its coordination shell. 
In the liquid state, time averaging over every atom is equal to the ensemble average. 
In the glass state, as the coordination numbers of many atoms are fixed on the timescale
of a simulation, time average of the stress for every particular atom is not equal to the ensemble
average. Thus, breakdown of equipartition of the atomic level stress energies
at the glass transition might signal ergodicity breaking. 

Finally, we note that the total potential energy of the system
per atom follows $(3/2)k_bT$ law (see Fig.3 of Ref.\cite{Levashov2008E})
in the glass state, i.e., it follows the equipartition law 
and it grows faster than the energies of the atomic level stresses.
Thus, the energies of the atomic level stresses obtained from our calculations here
or in the previous MD simulations do not capture the total increase
in potential energy.

In the liquid state, the total 
potential energy grows faster than $(3/2)k_bT$ in the range
of temperatures between the glass transition temperature and 
the potential energy landscape crossover temperature 
(see Fig.3 of Ref.\cite{Levashov2008E}). Thus, the total potential energy 
again grows faster than the energy of the atomic level stresses.
In this range of temperatures, the total potential energy follows 
the Rosenfeld-Tarazona law $U=U_g + bT^{3/5}$ 
\cite{Rosenfeld1998,Gebremichael2005}, 
where $U_g$ is the energy at the glass transition.   
The value of the coefficient $b$ is such that in the 
studied range of temperatures (which covers all reasonable temperatures),
potential energy of a liquid is larger than $(3/2)k_bT$.
Thus, the energies of the local atomic level stresses and the harmonic approximation
made in their derivations do not reflect
all processes that happen in the model liquid upon heating.

\section{Acknowledgments} 

We would like to thank T. Egami, V.N. Novikov, and K.A. Lokshin for useful discussions.


\begin{thebibliography}{10}
\bibitem{numerically}
For transverse waves it is convenient to represent $\bm{\hat{e}}_{\bm{q}}^T$
as a linear combination of vectors 
$\bm{\hat{e}}_{\bm{q}}^1$, $\bm{\hat{e}}_{\bm{q}}^2$ 
which are orthogonal to $\bm{\hat{q}}$ and to each other. 
If angles $\theta$ and $\phi$ define the direction of $\bm{q}$ 
then we can chose
$\bm{\hat{e}}_{\bm{q}}^1 \equiv\left[\cos(\theta)\cos(\phi),\cos(\theta)\sin(\phi),-\sin(\theta)\right]$ and $\bm{\hat{e}}_{\bm{q}}^2 \equiv \left[ \bm{\hat{q}} \times \bm{\hat{e}}_{\bm{q}}^1\right]=
\left[-\sin(\phi),\cos(\phi),0\right]$. 
Thus we write $\bm{\hat{e}}_{\bm{q}}^T \equiv \bm{\hat{e}}_{\bm{q}}^1 \cos(\psi) + \bm{\hat{e}}_{\bm{q}}^2 \sin(\psi)$. In (\ref{eq;Sepsilon03}), 
after taking the square, 
averaging should be done over $\psi$ and then 
over the directions of $\bm{\hat{q}}$.
For longitudinal waves $\bm{\hat{e}}_{\bm{q}}^L || \bm{\hat{q}}$.
\end{thebibliography}
\end{document}